\documentclass[12pt]{article}
\usepackage{graphicx}
\usepackage{amssymb,mathtools}
\usepackage{amsmath}
\usepackage{color}
\usepackage{tabularx}
\usepackage{enumitem}
\usepackage{hyperref}
\usepackage{dsfont}
\usepackage{algorithm}
\usepackage{algpseudocode}
\usepackage{multirow}
\usepackage{threeparttable}
\usepackage{lscape}
\usepackage{booktabs}
\usepackage{setspace}
\usepackage{url}
\usepackage{caption2}

\newcommand{\blind}{1}

\usepackage[top=0.6in, bottom=0.7in,left=1in,right=1in]{geometry}
\usepackage{natbib}
 \bibpunct[, ]{(}{)}{,}{a}{}{,}%

\makeatletter
\newenvironment{breakablealgorithm}
  {
   \begin{center}
     \refstepcounter{algorithm}
     \hrule height.8pt depth0pt \kern2pt
     \renewcommand{\caption}[2][\relax]{
       {\raggedright\textbf{\ALG@name~\thealgorithm} ##2\par}%
       \ifx\relax##1\relax 
         \addcontentsline{loa}{algorithm}{\protect\numberline{\thealgorithm}##2}%
       \else 
         \addcontentsline{loa}{algorithm}{\protect\numberline{\thealgorithm}##1}%
       \fi
       \kern2pt\hrule\kern2pt
     }
  }{
     \kern2pt\hrule\relax
   \end{center}
  }
\makeatother

\newcommand{\abs}[1]{\left\lvert#1\right\rvert}
\newcommand{\norm}[1]{\left\lVert#1\right\rVert}

\global\long\def\real{\mathbb{R}}

\global\long\def\prob{\mathbb{P}}
\global\long\def\mean {\mathbb{E}}

\newtheorem{prop}{Proposition}

\begin{document}

\def\spacingset#1{\renewcommand{\baselinestretch}%
{#1}\small\normalsize} \spacingset{1}


\if1\blind
{
  \title{\bf Fast Signal Region Detection with Application to Whole Genome Association Studies}
  \author{Wei Zhang\\
    School of Mathematical Sciences, Center for Statistical Science, \\Peking University, Beijing, China\\
    Fan Wang \\
    Department of Biostatistics, Mailman School of Public Health, \\Columbia University, New York, U.S.A\\
    Fang Yao\thanks{
     Wei Zhang and Fan Wang are co-first authors, and Fang Yao is the corresponding author. E-mail: fyao@math.pku.edu.cn. This research has been conducted using the UK Biobank Resource under project 79237.\hspace{.2cm}} \\
    School of Mathematical Sciences, Center for Statistical Science, \\Peking University, Beijing, China}
  \maketitle
} \fi

\if0\blind
{
  \bigskip
  \bigskip
  \bigskip
  \begin{center}
    {\LARGE\bf Fast Signal Region Detection with Application to Whole-Genome Association Studies}
\end{center}
  \medskip
} \fi

\begin{abstract}

Research on the localization of the genetic basis associated with diseases or traits has been widely conducted in the last few decades. Scan methods have been developed for region-based analysis in whole-genome association studies, helping us better understand how genetics influences human diseases or traits, especially when the aggregated effects of multiple causal variants are present. In this paper, we propose a fast and effective algorithm coupling with high-dimensional test for simultaneously detecting multiple signal regions, which is distinct from existing methods using scan or knockoff statistics. The idea is to conduct binary splitting with re-search and arrangement based on a sequence of dynamic critical values to increase detection accuracy and reduce computation.  Theoretical and empirical studies demonstrate that our approach enjoys favorable theoretical guarantees with fewer restrictions and exhibits superior numerical performance with faster computation. Utilizing the UK Biobank data to identify the genetic regions related to breast cancer, we confirm previous findings and meanwhile, identify a number of new regions that suggest strong associations with risk of breast cancer and deserve further investigation. 
\end{abstract}

\noindent%
\vfill

\newpage 
\spacingset{1.9} 
\section{Introduction}
\label{sec:intro}

Identifying the genetic basis for human diseases or traits is a fundamental goal in human genetic research. Over the past fifteen years, whole-genome association studies have uncovered numerous signal regions, providing insights into the genetic architecture of complex diseases and traits. 
Early research on whole-genome association studies has focused on single-nucleotide polymorphism (SNP)-based analysis,  where the statistical significance of SNPs is assessed through appropriate models such as the generalized linear model \citep{cameron2013regression}. Selection of significant SNPs is then carried out by Bonferroni correction to control the family-wise error rate, e.g., REGENIE proposed by \cite{mbatchou2021computationally}, or through a variety of false discovery rate control procedures, e.g., the Benjamin-Hochberg procedure \citep{Benjamin1995} and the Knockoff method \citep{Candes2015,Candes2021}. Due to the polygenic nature of complex human diseases or traits, the effect size of a single SNP can be too small to be detected by SNP-based analysis,  even with a large sample size \citep{liu2010versatile,bakshi2016fast}.  To address this limitation, set-based association approaches have been developed to test the aggregated effects of multiple SNPs within a set, which are powerful when multiple causal variants are present in a region and have been widely adopted for the identification of rare variants in genetic studies. There is recent literature on these methods which jointly test the effects of multiple variants in a genomic region, such as burden tests \citep{MORGENTHALER200728, Madsen2009}, nonburden tests \citep{Lin2011AJHG, Lee2012Biostatistics} and Sequence Kernel Association Test (SKAT) \citep{Wu2011AJHG}.  However, traditional set-based approaches require prespecifying the genetic regions (e.g., genes), which is not applicable for intergenic regions of the genome where the analysis units are not well-defined. The 4S algorithm proposed in \cite{Ning2021SSSS} combines individual and set-based analysis, first identifying significant points and then detecting signal regions around them. However, the 4S algorithm assumes that features are mutually independent, which is an overly restrictive assumption for general signal region detection problems.

Statistical methods have been developed to identify the size and locations of signal regions without prespecifying regions. Specifically, \cite{Naus1982} proposed the scan statistic to search the human genome continuously with a prespecified window size, and then mean-based scan statistic procedures have been developed for DNA copy number analysis, allowing a collection of window sizes under the setting closely related to the change-point detection problems \citep{Cai10, Olshen04, Zhang10}. 
However, these procedures assuming that all observations have the same mean within signal regions and individual variants are independently and identically distributed generally do not hold in whole-genome association studies, given the presence of linkage disequilibrium between variants. 
 To address the aforementioned limitations, \cite{Lin20} proposed a quadratic scan statistic for aggregating information in certain intervals, and combined it with the scan algorithm presented in \cite{Cai10} to detect signal regions at genome scale. This procedure has been demonstrated to outperform other existing methods, especially when the signals are in different directions, or when signal regions contain both causal and neutral variants. However, the scan-based methods face two major challenges. At first, most scan-based methods detect the whole genome region continuously and pick those non-overlapped regions significant enough in the sense that the related statistics are larger than a certain threshold, which is computationally intensive and tends to be conservative.  Secondly, when the signal strengths of different regions are severely unbalanced, a fixed threshold may miss some regions with relatively weaker signals.  Other recently developed approaches include those based on the knockoff framework \citep{he2021identification, he2021genome}.  For example,  the KnockoffScreen \citep{he2021identification} conducts window-based analysis to scan the whole genome and can help distinguish signals due to correlation confounding.  However,  generating knockoff features for each SNP requires more computation than other existing approaches. 

\begin{figure}[htbp]
\centering
\includegraphics[width=14cm]{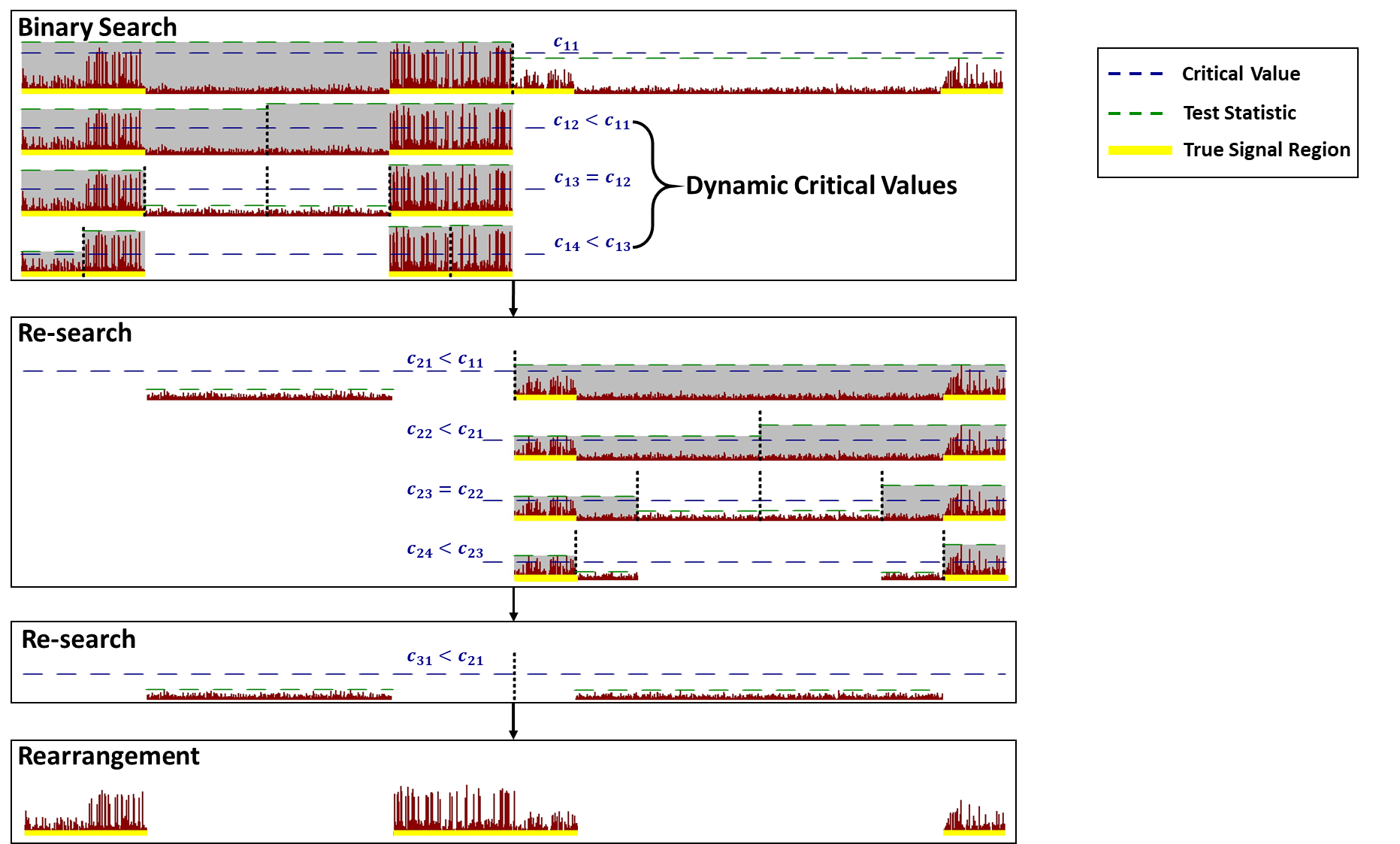}
\caption{BiRS detection scheme with extreme-type test. The red lines are observed signals and the grey parts are significant regions.}
\label{fig:scheme}
\end{figure}

By thoroughly surveying the literature in the field, we identify that designing a faster localization algorithm coupled with suitable/powerful tests, which overall has guarantees on both size control and detection power, is of fundamental importance and poses a primary challenge for signal region detection in whole-genome association studies. In this paper, we propose a new algorithm, referred to as the binary and re-search (BiRS) algorithm illustrated in Figure \ref{fig:scheme}. Instead of continuously scanning the entire region, it detects signal regions by iteratively splitting identified regions until the minimum size is reached. This binary search enables the fast identification of signal regions without prior knowledge of the starts and ends of the segments. Then a re-search procedure is applied to remove significant regions and initiates a new search among the remaining regions. We carefully employed a sequence of dynamic critical values that are readjusted during each step of testing to increase the power. We emphasize that the re-search procedure is essential to achieve high detection accuracy, as a true signal region may be divided into two or more segments, and the insignificant segments can be reassessed by readjusted critical values. 
We continue applying the re-search procedure until no more significant regions are identified. Finally, we re-arrange all selected regions and combine adjacent ones, thereby automatically adapting to the underlying signal structure.

To the best of our knowledge, this is the first attempt beyond scan-based algorithms to design a fast algorithm that effectively combines a suitable high-dimensional test and dynamic critical values that are not only powerful but also facilitate fast computation, which departs from conventional algorithms that combine with low-dimensional tests in genetic association studies. The BiRS algorithm offers three main advantages. At first, it exhibits significantly higher detection power than state-of-the-art signal region detection approaches such as scan and knockoff-based methods, while effectively controlling the family-wise error rate (FWER) through the construction of a sequence of dynamic critical values. Through re-search procedures, it also enables us to identify shorter or longer regions with unbalanced signal strengths, accommodating more dependence structures. Secondly, the BiRS algorithm enables significantly faster computation by iteratively binary splitting regions, with computational complexity allowing it to compute hundreds of times faster than the scan algorithm at genome scale. Thirdly, the BiRS algorithm enjoys favorable theoretical guarantees and fewer restrictions compared to Q-SCAN, and is flexible to be combined with most high-dimensional tests, whereas scan algorithms lack the capability to be combined with extreme-type tests.\footnote{For example: let $T = (1, 2, 3, 0, 0)'$, and the test statistic in region $I$ is $T_I = \max_{i\in I}T_i$. Let $I_1 = \left\{1, 2, 3\right\}$ and $I_2 = \left\{3, 4, 5\right\}$, then $T_{I_1} = T_{I_2}$.} Here we couple the BiRS algorithm with the distribution and correlation-free (DCF) test proposed in \cite{Xue20}, which allows fast computation at the vector level and utilizes a multiplier bootstrap to obtain accurate critical values.

We apply the BiRS algorithm to analyze genetic data from the UK Biobank, aiming to identify signal regions for breast cancer. Our genome-wide association study (GWAS) successfully validates several previously known SNPs, such as rs3817198 in the LSP1 gene \citep{nourolahzadeh2020correlation,afzaljavan2020predictive}, while also identifying novel signal regions on chromosomes 19 and 12, which encompass the genes \textsl{KCNN4}, \textsl{LYPD5}, and \textsl{CCDC91}. These regions, along with newly detected SNPs in \textsl{LINC01488}, indicate significant associations with breast cancer risk and its underlying regulatory pathways. We further demonstrate the broader applicability of our method by analyzing both coding and non-coding regions of the genome using the UK Biobank 200K whole-genome sequencing (WGS) dataset. Our analysis identifies rare variants in susceptibility genes that are also detected in our GWAS, such as \textsl{NEK10}, \textsl{SLC4A7}, \textsl{MRPS30}, \textsl{CCDC170}, and \textsl{ESR1}. Additionally, we uncover numerous rare variants within these genes that have been underpowered in earlier studies. Moreover, novel rare variants are identified in the non-coding regions of six genes, including \textsl{MAP3K1}, \textsl{SETD9}, and \textsl{MIER3}, shedding light on previously uncharacterized regulatory mechanisms. These findings demonstrate the usefulness of BiRS-DCF in modern WGS studies, opening new avenues for understanding the genetic architecture of breast cancer.

The rest of the article is organized as follows. In Section \ref{sec:algorithm}, we briefly review the DCF test and describe the proposed BiRS algorithm in detail. We conduct comprehensive simulations in Section \ref{sec:simulation} to demonstrate that the proposed method enjoys preferably numerical performance compared with existing approaches. In Section \ref{sec:realdata}, we apply the BiRS algorithm to conduct GWASs and WGS studies on UK Biobank data, aiming to identify signal regions for breast cancer. We conclude the article with a discussion in Section \ref{sec:dis}. Theoretical properties, technical assumptions, lemmas, and proofs of the theoretical results, along with more discussions on various issues, additional simulation and application results, are provided in the Supplementary Material for space economy. Lastly, we make available the R-package {\tt BiRS} of the simulation and application codes, along with the results from the GWAS and WGS study at \url{https://github.com/ZWCR7/Supplementary-Material-for-BiRS.git}.

\section{Proposed Method: Binary and Re-Search (BiRS)}
\label{sec:algorithm}
In this section, we describe the BiRS algorithm integrating the DCF two-sample mean test \citep{Xue20}. For ease of exposition, we first briefly review the DCF test and then describe how the BiRS algorithm is conducted. Further discussions about the choice of tests are provided in Supplementary Material.

\subsection{Review of DCF two-sample mean test}
\label{subsec:DCF}
Denote two independent genotype matrices by $\boldsymbol{X} = \left( X_1, \dots, X_n \right)^T$ and $\boldsymbol{Y} = \left( Y_1, \dots ,Y_m \right)^T$, respectively, where $X_1, \dots, X_n \in \real^p$ are genotypes for individuals that are mutually independent but not necessarily identically distributed with $\mean\left(X_i\right) = \mu^{X} = \left(\mu_1^X, \dots, \mu_p^X\right)^T$, $i = 1, \dots, n$, and $Y_1, \dots, Y_m$ are defined similarly with $\mean\left(Y_i\right) = \mu^{Y} = \left( \mu_1^Y, \dots, \mu_p^Y \right)^T$, $i = 1, \dots, m$. The hypothesis of interest is,
\vspace{-0.1in}
\begin{equation}
\vspace{-0.1in}
H_0:\mu^X = \mu^Y \quad\mathrm{v.s.}\quad H_1:\mu^X \neq \mu^Y. \label{test}
\end{equation}
Define the normalized sums $S_n^X = n^{-1/2}\sum_{i=1}^nX_i$ and $S_m^Y = m^{-1/2}\sum_{i=1}^mY_i$, respectively. Then the proposed DCF test statistic is given by
$T_{n,m} = \norm{S_n^X - n^{1/2}m^{-1/2}S_m^Y}_{\infty},$
where $\|a \|_\infty=\max\{\abs{a_1}, \ldots, \abs{a_p}\}$ for $a\in \real^p$.
This test will reject the null hypothesis at a certain significance level $\alpha$ if $T_{n,m} > c(\alpha)$ for some critical value $c(\alpha)$. The derivation of $c(\alpha)$ follows from the high-dimensional central limit theorem in \cite{chernozhukov2017central}. Precisely, let $\bar{X} = n^{-1} \sum_{i=1}^n X_i$ and $\hat{\Sigma}^X = n^{-1} \sum_{i=1}^n\left( X_i - \bar{X} \right)\left( X_i - \bar{X} \right)$ be the sample mean and sample covariance matrix of $X$, respectively, and define $\bar{Y}$ and $\hat{\Sigma}^Y$ analogously. Let $\boldsymbol{e} = \left\{ e_i \right\}_{i=1}^{n+m}$ represent a set of independently and identically distributed (i.i.d.) standard normal random variables, and define 
\vspace{-0.1in}
\begin{equation} 
\vspace{-0.1in}
\label{Sboot}
S_n^{eX} = \frac{1}{\sqrt{n}}\sum_{i=1}^n e_i\left( X_{i} - \bar{X} \right); S_m^{eY} = \frac{1}{\sqrt{m}}\sum_{i=1}^m e_{i+n}\left( Y_{i} - \bar{Y} \right),
\end{equation}
Then $c(\alpha)$ is calculated by
\vspace{-0.1in}
\begin{equation}
\vspace{-0.1in}
\label{critical}
 c(\alpha) = \inf \left\{ t \in\real : \prob_e\left( \norm{S_n^{eX} - n^{1/2}m^{-1/2}S_m^{eY}}_{\infty} \leq t \right) \geq 1 - \alpha \right\},
\end{equation} 
where $\prob_e\left( \cdot \right)$ denotes the probability measure with respect to $\boldsymbol{e}$ only. In practice, this critical value can be approximated by the Gaussian multiplier bootstrap\citep{chernozhukov2013gaussian}. For ease of presentation, we present the DCF bootstrap testing procedure below.

\begin{breakablealgorithm}
    \setstretch{1}
    \renewcommand{\algorithmicrequire}{\textbf{Input:}}
    \renewcommand{\algorithmicensure}{\textbf{Output:}}
    \caption{DCF test procedure}\label{alg:DCF}
    \begin{algorithmic}[1]
    \State\textbf{Input}: genotype matrices $\boldsymbol{X}, \boldsymbol{Y}$, number of bootstraps $N$.
    \hspace*{0.02in}
    \State Calculate the test statistics 
    $$ T_{n,m} = \norm{S_n^X - n^{1/2}m^{-1/2}S_m^Y}_{\infty}. $$
    \State Generate $N$ sets of standard normal random variables with size $(n + m)$ of each, denoted by $\boldsymbol{e}_1, \dots, \boldsymbol{e}_N$ as random copies of $\boldsymbol{e}$.
    \State Calculate $N$ times of $T_{n,m}^e = \norm{S_n^{eX} - n^{1/2}m^{-1/2}S_m^{eY}}_{\infty}$, denoted as $\left\{ T_{n,m}^{e_1}, \dots, T_{n,m}^{e_N} \right\}$.
    \State Take $\hat{c}(\alpha)$ as the $100(1 - \alpha)$th sample percentile of $\left\{ T_{n,m}^{e_1}, \dots, T_{n,m}^{e_N} \right\}$ to approximate $c(\alpha)$.
    \If{$T_{n,m} > \hat{c}(\alpha)$}
       \State\Return 1 (The DCF test rejects the null hypothesis).
    \Else
       \State\Return 0 (The DCF test accepts the null hypothesis).
    \EndIf
    \end{algorithmic}  
\end{breakablealgorithm}

Since $\hat{c}(\alpha)$ tends to $c(\alpha)$ when the number of bootstraps $N \rightarrow \infty$, hereafter, without loss of generality, we use the population version $c(\alpha)$ rather than the empirical version $\hat{c}(\alpha)$ for illustration.


\subsection{Binary splitting}

\label{subsec:bisr}

We introduce some key concepts used throughout the paper. Under the alternative hypothesis of \eqref{test}, a signal point is defined at index $j$ if $\mu_j^X \neq \mu_j^Y$, $1 \leq j \leq p$, and let $I$ be an index set containing certain signal points. We call $I$ a signal region if it is continuous and separable. Specifically, a region $I$ is said to be continuous if $\mu_I^X \neq \mu_I^Y$ , where the `$\neq$' means that for any $j \in I$, $\mu_j^X \neq \mu_j^Y$. Let $r_1 = \min\left\{j: j \in I\right\}$ and $r_2 = \max\left\{j: j \in I \right\}$. Then $I$ is said to be seperable if  $\mu_{r_1-1}^X = \mu_{r_1-1}^Y$ and $ \mu_{r_2+1}^X = \mu_{r_2+1}^Y$. Lastly, we denote the true signal regions by $I_1^*, \dots, I_\ell^*$ (if exist) and let $\mathcal{I}^* = \left\{ I_1^*, \dots, I_\ell^* \right\}$.

We are interested in testing whether signal regions exist, and if so, then identifying the locations of these signal regions. Hence, we first apply the DCF test for:
\vspace{-0.1in}
\begin{equation}
\vspace{-0.1in}
\label{global}
H_0: \mathcal{I} = \emptyset \quad\mathrm{v.s.}\quad H_1: \mathcal{I} \neq \emptyset,
\end{equation}
which is equivalent to the hypothesis testing in \eqref{test}. If the global null hypothesis cannot be rejected, i.e., $T_{n,m} \leq c(\alpha)$, we conclude that no signal regions exist and terminate the search. If $T_{n,m} > c(\alpha)$, we continue the binary search to find the specific locations of the signal regions as follows.



We split the genotype matrices into two parts and conduct a two-sample DCF test for each half. Without loss of generality, let $I_{11} = \left\{1, 2, \dots, \lfloor p/2 \rfloor\right\}$ and $I_{12} = \left\{\lfloor p/2 \rfloor + 1, \dots, p\right\}$ with $\lfloor \cdot\rfloor$ taking value of the  largest integer below. Let $\boldsymbol{X}_{I_{11}}, \boldsymbol{Y}_{I_{12}}$ contain the columns of $\boldsymbol{X}, \boldsymbol{Y}$ in region $I_{11}$ ($\boldsymbol{X}_{I_{12}}, \boldsymbol{Y}_{I_{12}}$ for region $I_{12}$ similarly), and write $\boldsymbol{X}_1 = \left(\boldsymbol{X}_{I_{11}}, \boldsymbol{X}_{I_{12}}\right)$ and $\boldsymbol{Y}_1 = \left(\boldsymbol{Y}_{I_{11}}, \boldsymbol{Y}_{I_{12}}\right)$. We compute the DCF test statistics $T_{11}, T_{12}$ based on $\boldsymbol{X}_{I_{11}}, \boldsymbol{Y}_{I_{11}}$ and $\boldsymbol{X}_{I_{12}}, \boldsymbol{Y}_{I_{12}}$, i.e., $T_{11} = \norm{S_n^{X_{I_{11}}} - n^{1/2}m^{-1/2}S_m^{Y_{I_{11}}}}_{\infty}$; $T_{12} = \norm{S_n^{X_{I_{12}}} - n^{1/2}m^{-1/2}S_m^{Y_{I_{12}}}}_{\infty}$, 
where $S_n^{X_{I_{1k}}}, S_m^{Y_{I_{1k}}}, k = 1, 2$ are the normalized sums based on the sample matrices $\boldsymbol{X}_{I_{1k}}, \boldsymbol{Y}_{I_{1k}}$, $k = 1, 2$. Following Eq.\eqref{Sboot} and \eqref{critical}, the critical value $c_{1}(\alpha)$ is computed through the multiplier bootstrap based on the sample matrices $\boldsymbol{X}_1, \boldsymbol{Y}_1$ with independently generated $\boldsymbol{e}$ of standard normal random variables. Specifically,
\vspace{-0.1in}
\begin{equation*}
\vspace{-0.1in}
c_{1}(\alpha) = \inf\left\{ t \in \real: \prob_e\left( \norm{S_n^{eX_1} - n^{1/2}m^{-1/2}S_m^{eY_1}}_{\infty} \leq t \right) \geq 1 - \alpha \right\},
\end{equation*}
where $S_n^{eX_1} = \frac{1}{\sqrt{n}}\sum_{i=1}^n e_i\left( X_{1_i} - \bar{X}_1 \right)$; $S_m^{eY_1} = \frac{1}{\sqrt{m}}\sum_{i=1}^m e_{i+n}\left( Y_{1_i} - \bar{Y}_1 \right)$,and $X_{1_i}, Y_{1_i}$ are the $i$-th samples of genotype matrices $\boldsymbol{X}_1, \boldsymbol{Y}_1$, respectively. The subsequent $c_j(\alpha)$'s are obtained in a similar manner.

Regarding the termination of the binary search, we set a truncation parameter $s$ associated with the length of signal regions which is chosen empirically according to the detection problem at hand. For the first binary search with $k = 1, 2$, if $T_{1k} > c_{1}(\alpha)$ and $\abs{I_{1k}} > 2^s$, we take $I_{1k}$ as the possible signal region for next binary search; if $T_{1k} > c_{1}(\alpha)$ but $\abs{I_{1k}} \leq 2^s$(i.e., the region length falls below the pre-specified value) , we stop the binary splitting for region $I_{1k}$ and take $I_{1k}$ as one of the estimated signal segments. Otherwise, we conclude that there is no signal within region $I_{1k}$, and therefore stop the binary splitting for region $I_{1k}$. 

For the $j$-th binary search, where there exist $n_j, 1 \leq n_j \leq 2^j$ possible signal segments $I_{j1}, \dots, I_{jn_j}$, the corresponding genotype matrices are $\boldsymbol{X}_{I_{j1}}, \dots, \boldsymbol{X}_{I_{jn_j}}$ and $\boldsymbol{Y}_{I_{j1}}, \dots, \boldsymbol{Y}_{I_{jn_j}}$. Let $\boldsymbol{X}_j = \left( \boldsymbol{X}_{I_{j1}}, \dots, \boldsymbol{X}_{I_{jn_j}} \right)$ and $\boldsymbol{Y}_j = \left( \boldsymbol{Y}_{I_{j1}}, \dots, \boldsymbol{Y}_{I_{jn_j}} \right)$. Then the DCF test statistics $T_{jk}, k = 1, \dots, n_j$ and critical value $c_{j}(\alpha)$ are obtained in the same manner. Analogously, for $k = 1, \dots, n_j$, when $T_{jk} > c_{j}(\alpha)$, we take $I_{jk}$ as the possible signal segment, either for the next search if $\abs{I_{jk}} > 2^s$ or stop the splitting if $\abs{I_{jk}} \leq 2^s$; otherwise, we regard that no signal within region $I_{jk}$. If there is not any signal region to be further divided, then the first round of binary search will be terminated with the detected signal segments denoted by $\hat{I}^{(0)}_{1}, \dots, \hat{I}^{(0)}_{K_0}$, where the superscript $^{(0)}$ stands for this initial search. We remark that in this binary search, the dimensions of the sample matrices decrease, which results in dynamic critical values $c_{j}(\alpha)$ and guarantees the improvement of testing power during the whole procedure.

In this subsection, we propose a binary search algorithm that utilizes a sequence of carefully chosen dynamic critical values. Unlike the use of a single global threshold in most existing work, these dynamic critical values allow us to detect more signals. Specifically, since $\boldsymbol{X}_j$, $\boldsymbol{Y}_j$ are columns of the matrices $\boldsymbol{X}$ and $\boldsymbol{Y}$, respectively. Let the dimensions of $\boldsymbol{X}_j, \boldsymbol{Y}_j$ be $p_j$,  According to the theoretical results in the proof of Lemma S4, we can derive that $c(\alpha)$ is of the order $\sqrt{B_{n,m}\log(pn)}$ and consequently, $c_j(\alpha) \lesssim \sqrt{B_{n,m}\log(p_jn)}$. Since signals are usually sparse in practice, $p_j \ll p$ for a large $j$. This binary splitting significantly reduces the dimensionality, leading to a substantial reduction in the dynamic critical values, which in turn helps us detect more signals.

Regarding computation, recall that $S_n^{eX_j} - n^{1/2}m^{-1/2}S_n^{eY_j}$ is a sub-vector of  the vector $S_n^{eX} - n^{1/2}m^{-1/2}S_n^{eY}$. To obtain the critical value, one only needs to calculate the infinity norm restricted to this sub-vector, which is computationally efficient.

\subsection{Re-search with rearrangement}
\label{subsec:re-search}

When signal strengths are unbalanced, almost all tests tend to detect the strong signals and miss the weak ones, which results in loss of power and is more severe in the high dimensional tests. For instance, the scan algorithm detects signals by selecting the regions whose estimated signal strengths are stronger than a common threshold and misses the regions with weaker signals. In the proposed binary search, the use of the dynamic critical values $c_{j}(\alpha)$ can alleviate this problem to a certain degree. However, as we use the same $c_{j}(\alpha)$ for all tests of $I_{j_1},\ldots, I_{j_{n_j}}$, there still possibly exist some signals missing within each step if they are unbalanced (see Example S1 in Supplementary Material for details). 
To address this issue, we propose a re-search algorithm as the second phase of our BiRS algorithm, which proceeds as follows.

Recall that the detected signal segments are $\hat{I}_{1}^{(0)}, \dots, \hat{I}_{K_0}^{(0)}$ and let $\hat{I}^{(0)} = \cup_{j=1}^{K_0} \hat{I}_j^{(0)}$. We substitute the variables of segment $\hat{I}^{(0)}$ in sample matrices $\boldsymbol{X}, \boldsymbol{Y}$ with zeros, denoted by $\boldsymbol{X}^{(1)}, \boldsymbol{Y}^{(1)}$. 
This operation aims to detect weaker signals that may have been missed. Recall that the dynamic critical values are $1 - \alpha$ quantiles of the infinity-norm of $S_n^{eX} - n^{1/2}m^{-1/2}S_n^{eY}$ or its sub-vectors, replacing certain columns of sample matrices reduces the values of infinity norms, and then reduces the dynamic critical values, which consequently facilitates the detection. Although the same reductions can be obtained by removing selected segments, this operation preserves the locations and lengths of the candidate intervals during the whole procedure while the removing operation does not, which facilitates computation and theoretical analysis.

Then we conduct the global DCF test using sample matrices $\boldsymbol{X}^{(1)}, \boldsymbol{Y}^{(1)}$. If the global test does not reject the null hypothesis, we stop the re-search algorithm. Otherwise, we apply the binary search for $\boldsymbol{X}^{(1)}, \boldsymbol{Y}^{(1)}$, and denote the detected signal segments as $\hat{I}^{(1)}_{1}, \dots, \hat{I}^{(1)}_{K_1}$ and $\hat{I}^{(1)} = \cup_{j = 1}^{K_1} \hat{I}^{(1)}_{j}$. For $l>1$, we repeat the re-search algorithm on $\boldsymbol{X}^{(l)}, \boldsymbol{Y}^{(l)}$ that are obtained by substituting the variables of sample matrices $\boldsymbol{X}^{(l-1)}, \boldsymbol{Y}^{(l-1)}$ in region $\hat{I}^{(l-1)}$ with zeros. The re-search algorithm stops at step $M$ if the global DCF test with respect to sample matrices $\boldsymbol{X}^{(M)}, \boldsymbol{Y}^{(M)}$ do not reject the null hypothesis.

Finally, note that the detected signal segments $\hat{I}^{(0)}_1, \dots, \hat{I}^{(0)}_{K_0}, \dots, \hat{I}^{(M)}_1, \dots, \hat{I}^{(M)}_{K_M}$ are continuous, and some of them may be neighboring. Here we say the continuous segments $I_1, I_2$ are neighboring if the union of them, i.e., $I_1 \cup I_2$ is also continuous. To obtain the signal regions satisfying separability, we rearrange the detected segments after the binary and re-search procedure, i.e., combining the neighboring segments, which gives the final estimate of signal regions denoted as $\hat{I}_1, \dots, \hat{I}_{\ell'}$. We stress that this procedure adapts to the structures of signal regions. For instance, a signal region has been chopped into several pieces during the binary splitting. Then all the pieces can be detected through the re-search procedure as long as the signal strength in each piece satisfies Assumption S7 in Supplementary Material and the underlying signal region will be recovered by rearrangement. The BiRS algorithm is summarized in Algorithm \ref{alg:BiRS} below, while a graphical illustration has been provided in Figure \ref{fig:scheme}.
\vspace{0.3in}

\begin{breakablealgorithm}
\setstretch{1}
\renewcommand{\algorithmicrequire}{\textbf{Input:}}
\renewcommand{\algorithmicensure}{\textbf{Output:}}
\caption{BiRS Algorithm for Signal Region Detection}\label{alg:BiRS}
\begin{algorithmic}[1]
\State\textbf{Input}: sample matrices $\boldsymbol{X}, \boldsymbol{Y}$, truncation parameter $s$.
\hspace*{0.02in}
\State Conduct global (DCF) test for the sample matrices $\boldsymbol{X}, \boldsymbol{Y}$.
\If{The test accepts the null hypothesis}
   \State \Return NULL (There is no signal region).
\Else
   \State Provide $s$, conduct the binary search based on $\boldsymbol{X}, \boldsymbol{Y}$.
   \State Replace the elements of $\boldsymbol{X}, \boldsymbol{Y}$ in estimated signal regions from last step as zeros.
   \State Conduct the global test for the new $\boldsymbol{X}, \boldsymbol{Y}$.
   \While{The test rejects the null hypothesis}
       \State Provide $s$, conduct the binary search based on $\boldsymbol{X}, \boldsymbol{Y}$.
       \State Replace the elements of $\boldsymbol{X}, \boldsymbol{Y}$ in estimated signal regions from last step as zeros.
       \State Conduct the global test for the new $\boldsymbol{X}, \boldsymbol{Y}$.
   \EndWhile
   \State Rearrange the estimated signal regions to be continuous and separated.
   \State\Return Estimated signal regions $\hat{I}_1, \dots, \hat{I}_{\ell'}$.
\EndIf
\end{algorithmic}  
\end{breakablealgorithm}

For the choice of truncation parameter $s$, we recommend to choose $s = [\log_2(4\alpha \hat{L}_{\min})]$ to control the false discovery rate at level $\alpha$, where $\hat{L}_{\min}$ is an approximate minimum length of signal regions and $[\cdot]$ is the rounding operator. More discussion and numerical analysis for the sensitivity of $s$ are in Section S2 in Supplementary Material. Additionally, we suggest to standardize the covariates before analysis, as the scales of covariates may influence the critical value $c(\alpha)$ that largely depends on the maximum eigenvalue of the covariance matrix. A scaling procedure can reduce the maximum eigenvalue and thereby help identify weaker signals in coordinates with smaller variances.

\subsection{Computational comparison with scan method}
\label{subsec:computational}
To demonstrate the computational advantage of the BiRS algorithm, we compare it with the commonly used scan method \citep{Lin20}, focusing on the total number of tests conducted in such detection procedures.

Recall that the estimated signal regions in scan algorithm are constructed by scanning all the continuous intervals with pre-specified lengths $ L_1> \dots >L_r$. Note that the number of candidate intervals with length $L_j$ is $(p - L_j)$, thus one should compute $\sum_{j=1}^r\left(p - L_j\right)$ test statistics for completing the scan search. Since the true signal regions are usually short, e.g., $o(p)$, then the number of test statistics is around $rp$ (or $2rp$ including the threshold calculation). In practice, $r$ is often moderately large and in most GWASs problems $r$ is around $50$.
For BiRS algorithm, a simple derivation indicates the following.
\begin{prop}
\label{prop:computational}
Assume that the BiRS algorithm stops after the $m$th re-search, then the number of test statistics needed does not exceed $(m + 1) \left(p/2^{s - 1} + \log_2 p  - s\right)$, where $s$ is the truncation parameter.
\end{prop}
According to the proof of Theorem S3 in Supplementary Material, if there are $\ell$ true signal regions, then $m \leq \ell$ with high probability. In practice, the true signal regions are usually sparse, e.g., $\ell=O(2^{s})$ in most detection problems.
In other words,  the BiRS algorithm needs to compute at most $(\ell + 2) p/2^{s - 1} \approx 2p$ test statistics, which is considerably less than the scan algorithm and empirically demonstrated in Section \ref{sec:simulation}.

\subsection{Theoretical guarantees}
For the flow of exposition, we present the theoretical analysis of the family-wise error rate and the detection accuracy in Section S6 of Supplementary Material. For size analysis, under mild conditions of the structure of the random vectors, the BiRS algorithm asymptotically controls the family-wise error rate of any multiple tests in the algorithm, and thus asymptotically controls the size of test problem \eqref{global}. The method allows the dimension to grow at an exponential rate relative to sample size and allows the truncation parameter $s$ and the number of true signal regions $\ell$ tend to infinity as $p \rightarrow \infty$. For detection accuracy analysis, we prove that the BiRS algorithm can consistently detect signal regions of more general structures than those by Q-SCAN. Specifically, the BiRS algorithm allows the signal strength in a certain region to decay at a rate depending on the whole region dimension minus the dimension of regions with strong signals, while the Q-SCAN needs the signal strengths across regions to be balanced. The proposed method relaxes the $M$-dependence assumption in Q-SCAN and the common scan-based method to be a ``weak'' dependence assumption, which allows a long-range correlation (LD) and is more applicable in genetic association studies. Moreover, the BiRS algorithm has less restriction on the lengths of signal regions and can consistently detect shorter or longer signal regions than Q-SCAN. See Theorem S1, S2 and S3 for more details.

\section{Simulation Studies}
\label{sec:simulation}

In this section, we conduct simulation studies to compare the proposed BiRS method with other region detection methods in terms of the true positive rate (TPR) and false discovery rate (FDR). For comprehensiveness, for the BiRS method, we combine it with the DCF test (BiRS-DCF) and the CL test (BiRS-CL) proposed by \cite{Cai14}, respectively. For scan-based methods, we consider the Q-SCAN method in \cite{Lin20} and the LRS procedure in \cite{Cai10} for comparison. Since the scan algorithm is not suitable for extreme-type tests as aforementioned,  we do not combine it with the DCF test or CL test. We also consider the KnockoffScreen method proposed in \cite{he2021identification} and the 4S algorithm in \cite{Ning2021SSSS} coupled with DCF test.

We compare the methods mentioned above in two different scenarios. The first setting is the {\it normal} setting which assumes that $X$ and $Y$ follow multivariate normal distributions, whereas the second is the {\it genetic} setting in which $X$ and $Y$ are simulated based on real genotype data which takes values of 0,1, or 2. Moreover, in genetic setting, we generate the signal regions based on the simulated functional annotations to better represent the intended application environment and add the SCANG-STAAR method in \cite{Li2022nature} 
for comparison. The detailed designs of these two settings are as follows.

\subsection{Normal Setting}
In the \textit{normal} setting, we first generate samples $\left\{ X_i \right\}_{i=1}^n$ and $ \left\{ Y_i \right\}_{i=1}^m$ which are i.i.d normal random variables, with sample sizes $n=600, m=400$ and dimension $p = 2^{13}=8192$. Without loss of generality, the mean of $Y$ is assumed to be $\mu^Y =0$. For $\mu^X$ (the mean of $X$), we set $\beta$ signal regions, denoted by $I_1,\ldots,I_\beta$, and consider the setting of $\beta=1,\ldots,4$. The signal regions $I_1,\ldots, I_4$ are generated as follows. (1) \textit{Lengths of signal regions}: sample 4 elements from $\mathcal{L} = \left\{ 128, 160, \cdots, 320 \right\}$ as the lengths of true signal regions which are denoted by $L_1, \dots, L_{4}$. (2) \textit{Locations of signal regions}: recall that the length of  the whole region is $p=8192$. Let $loc_j=(j/8)p$, $j=0,1,\ldots,7$ be the points which divide the whole region equally into eight parts and $\{k_1,\ldots,k_4\}$ be a permutation of $\{1,2,3,4\}$. We set $I_i=\{loc_{2i-1}+1,\ldots,loc_{2i-1}+L_{k_i}\}$, $i=1,2,3,4$ as the signal regions, which are non-overlapped. (3) \textit{Signal strengths}: for each signal region $I_j$ with length $L_{k_j}$, we generate $\gamma L_j$ random variables from the uniform distribution $U(-\delta, \delta)$ to mimic the sparse and strong signals and $(1-\gamma)L_j$ random variables from the uniform distribution $U(-\delta_0, \delta_0)$ to mimic the regular signals for normal setting, where $\delta>\delta_0$. 
 
We denote the covariance matrices of the random vectors as $\Sigma^{X} = \operatorname{Cov}(X_i), \Sigma^{Y}$ $ = \operatorname{Cov}(Y_{i'})$ for all $i = 1, \dots, n; i' = 1, \dots, m$. Under the normal setting,  we consider scenarios with two different covariance structures: an $M$-dependence structure and a weak-dependence structure.  For the $M$-dependence structure, we set the covariance matrix $(\Sigma)_{jk} = (1 + \abs{j - k})^{-1/4}\cdot\boldsymbol{1}_{\{\abs{j - k} \leq M\}}$ with $M = 64$.  For the weak dependence structure,  we set $(\Sigma)_{jk} = 0.90^{\abs{j - k}}$.  Under each $\Sigma$, we consider the following two scenarios: an equal covariance matrix such that $\Sigma^{X} = \Sigma^{Y} = \Sigma$, and unequal covariance matrices such that $\Sigma^{X} = 2\Sigma^{Y} = 2\Sigma$.

We set $\gamma = 0.25$, and begin with $\delta = 0.2$ and $\delta_0 = 0.05$, then strengthen the signals by setting $\delta = 0.25, 0.3, 0.35, 0.4$. To demonstrate the empirical FWER, we also set $\delta = \delta_0 = 0$. Since KnockoffScreen is an FDR-controlled procedure that is not designed for FWER control \citep{he2021identification}, thus we only compare BiRS-DCF, BiRS-CL, Q-SCAN, 4S and LRS for FWER. The global significant level $\alpha$ is 0.05. The BiRS-DCF method and the Q-SCAN method are conducted with the multiplier bootstrap of size $N = 1000$. The BiRS-CL test is constructed following \cite{Cai14}. We perform these methods based on $1000$ Monte Carlo runs over different $\Sigma^X, \Sigma^Y$ for computing FWERs and $500$ Monte Carlo runs over different $\beta$, $\delta$, $\Sigma^X,\Sigma^Y$ for computing FDRs and TPRs. The FWERs of all settings are presented in Table \ref{table:FWER}. The FDRs and TPRs of scenarios with $\beta=1,4$ are in Figure \ref{fig:simulation_normal}. The results of scenarios with $\beta = 2, 3$ can be found in Figure S3 of Supplementary Material, which also contains numerical values of FDR and TPR in Tables S1--S4 for all settings as the plots may not be sufficiently visible.

\begin{figure}[htbp]
\centering
\includegraphics[width = 16cm]{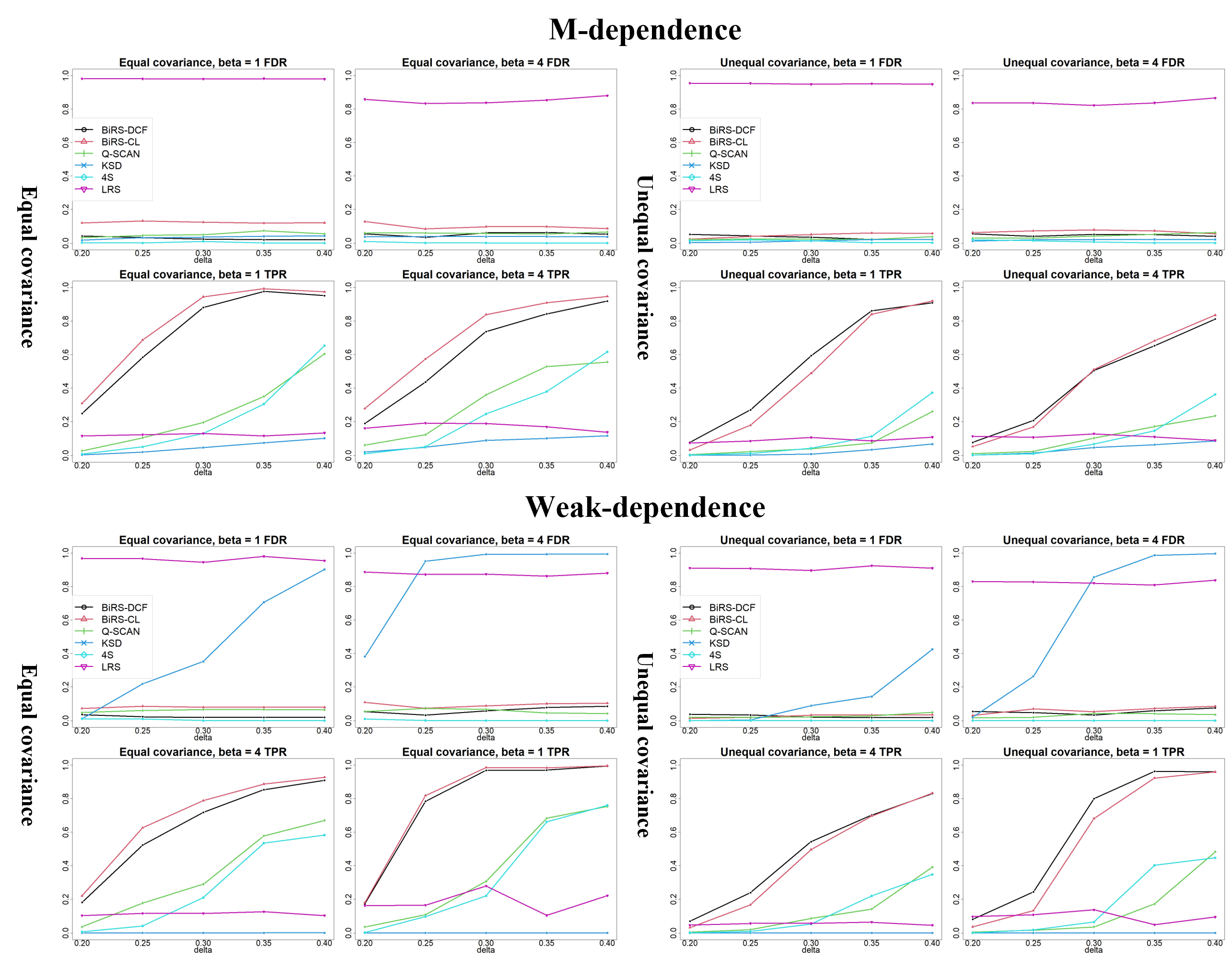}
\caption{\label{fig:simulation_normal}\small FDRs and TPRs versus signal strength $\delta$ of single signal regions and multiple signal regions under different covariance structures in normal setting. The top-left panel is the ``$M$-dependence with equal covariance setting"; the top-right panel is the ``$M$-dependence with unequal covariance setting"; the middle-left panel is the ``Weak-dependence with equal covariance setting"; the middle-right panel is the ``Weak-dependence with unequal covariance setting". The KSD in the legend refers to KnockoffScreen.}
\end{figure}

\begin{table}
  \setlength{\abovecaptionskip}{0cm}
  \caption{\label{table:FWER}\small FWERs of BiRS-DCF, BiRS-CL, Q-SCAN, LRS and 4S under different covariance structures in normal setting, the MES refers to ``$M$-dependence with equal covariance setting"; MNS refers to ``$M$-dependence with unequal covariance setting";  WES refers to ``Weak-dependence with equal covariance setting" and WNS refers to ``Weak-dependence with unequal covariance setting".} 
  \setlength{\tabcolsep}{6mm}
  \renewcommand\arraystretch{0.5}
  \centering
    \begin{tabular}{c|ccccc}
    \toprule
      & BiRS-DCF & BiRS-CL & Q-SCAN & LRS & 4S \\
    \midrule
    MES & 0.054 & 0.060 & 0.052 & 0.999 & 0.000\\
    MNS & 0.054 & 0.011 & 0.027 & 0.985 & 0.000\\
    WES & 0.050 & 0.034 & 0.047 & 0.997 & 0.000\\
    WNS & 0.045 & 0.007 & 0.020 & 0.934 & 0.000\\
    \bottomrule
    \end{tabular}
\end{table}

From these results, one sees that the BiRS-DCF controls the FWERs well in all settings, while the FWERs of BiRS-CL are very unstable and Q-SCAN tends to be conservative under unequal covariance structure and genetic settings. The LRS algorithm loses control of FWERs in all settings, and the 4S algorithm with DCF test attains zero FWERs because the 4S algorithm excludes isolated significant points. Despite achieving the highest TPRs, BiRS-CL exhibits inflation in FDRs when subjected to normal settings with equal covariance across different values of $\beta$ and $\delta$. A possible reason is that the test in \cite{Cai14} depends on the limiting distribution of Gaussian maxima requiring $p \rightarrow \infty$ and it can not control the false detection in relatively small regions. Q-SCAN, KnockoffScreen and 4S control the FDRs but have relatively lower TPRs. In contrast, BiRS-DCF achieves comparable power to BiRS-CL while maintaining FDR at around 0.05. Under unequal covariance settings with multiple signal regions, BiRS-CL exhibits poor FDR control and Q-SCAN has lower TPRs compared to other methods. Additionally, KnockoffScreen fails in weak dependence covariance structure since it assumes a block diagonal covariance structure \citep{he2021identification}. For LRS, it fails in all settings as the simple likelihood ratio test may not be suitable for general signal detection.

\begin{figure}
\centering
\includegraphics[width=14cm]{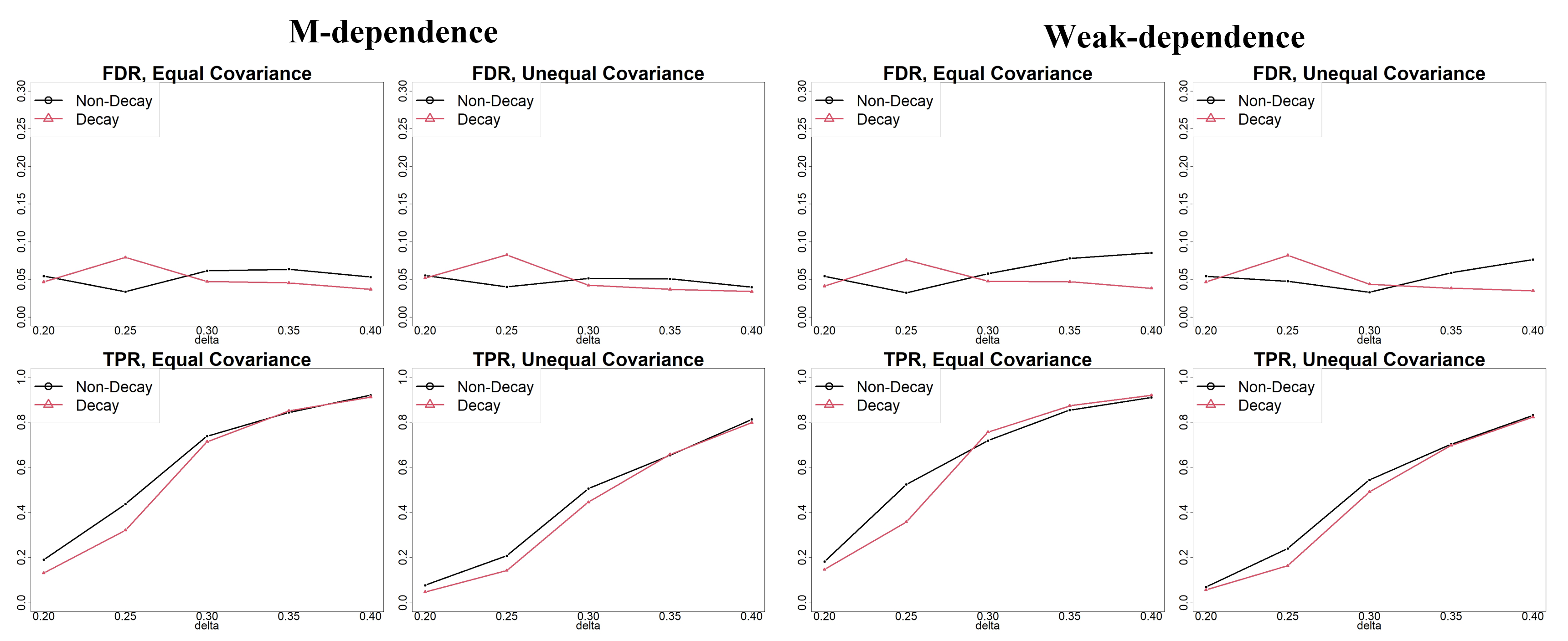}
\caption{\small Comparison of FDRs and TPRs of decay signal strength setting and non-decay signal strength. The left panel is $M$-dependence setting and the right panel is weak-dependence setting. The ``Decay'' refers to the decay setting and ``Non-decay" refers to the non-decay setting.}
\label{fig:decay}
\end{figure}

We also compute the FDRs and TPRs of BiRS-DCF method in the signal strength decay settings under normal setting. Let $\beta = 4$ and for $I_j, j \geq 2$, we subtract $\delta_1$ by $\rho_j=50 n^{-1/2} \Big\{\sqrt{\log(pn)} - \sqrt{\log\left[ \left( p - \sum_{k=1}^{j-1} L_j )\right]n \right)} \Big\}$. The detailed comparisons of TPRs and FDRs are provided in Figure \ref{fig:decay}, which indicates that BiRS-DCF has similar TPRs and FDRs in decayed and non-decayed signal settings and validates the theoretical finding in Theorem S2.

\subsection{Genetic Setting}

Under the \textit{genetic} setting, we first generate the sequence data in a 10-Mb region using the calibration coalescent model \citep{cosi2014} that mimics the linkage disequilibrium (LD) structure of European samples. Then we simulate the genotype matrices based on the LD matrix of the sequence data. Specifically, we randomly select 10 loci from the 10-Mb region with a total of 140096 SNPs and calculate the LD matrix by Pearson correlation coefficients among all SNPs. Following a common approach in genetic simulation studies \citep{Lai21}, we generate haplotypes for each individual from a correlated Bernoulli distribution based on the real LD matrix and obtain genotypes by combining the values in haplotypes. The sample sizes are chosen as $n = 6000$ and $m = 4000$.
 
For the mean difference, We randomly select four signal regions across the 10 loci. The lengths of the signal regions are randomly selected from lengths of 1kb, 1.5kb and 2kb. We consider the proportions of causal variants to be 10\% on average among the signal regions, and the probability of a variant being causal is allowed to depend on different sets of annotations through a logistic model, of which five are informative and the other five are noninformative. Specifically, for a point $j$ in a signal region, let $A_{j,1}, \dots, A_{j,10}$ be ten functional annotations that are i.i.d. from $\mathcal{N}(0,1)$ and $c_j$ denote whether point $j$ is causal. Then $\mathrm{logit}\mathbb{P}(c_j = 1) = \zeta_0 + \sum_{i=1}^5\zeta_{k_i}A_{j,k_i}$, where five annotations $\left\{ k_1, \dots, k_5 \right\} \subset \left\{1, \dots, 10\right\}$ are randomly sampled for the signal region. We set $\zeta_{k_i} = \log(4)$ and $\zeta_{0} = \mathrm{logit}(0.01)$, resulting in 10\% causal variants on average in signal regions.

\begin{figure}[htbp]
\centering
\includegraphics[width = 14cm]{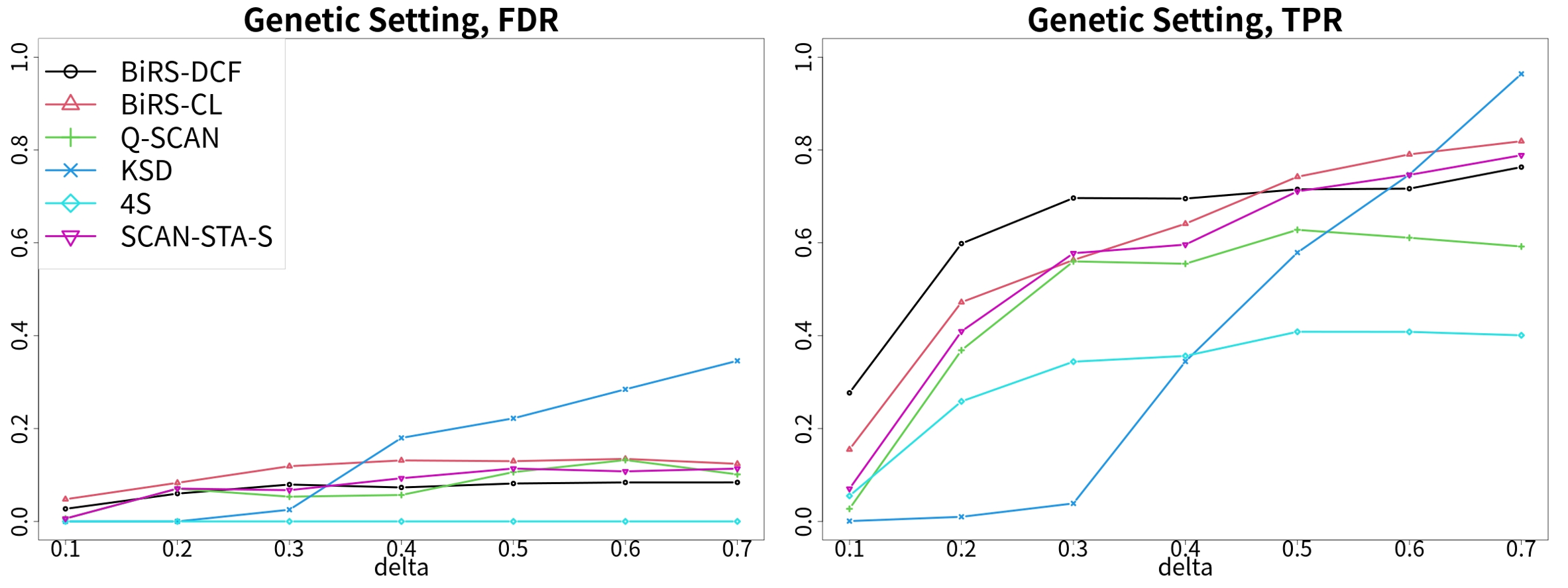}
\caption{\label{fig:simulation_genetic}\small FDRs and TPRs versus signal strength $\delta$ in genetic setting. The KSD in the legend refers to KnockoffScreen and SCAN-STA-S refers to SCANG-STAAR-S.}
\end{figure}

The signal strength at each signal point is drawn from $Sgn\cdot\delta\cdot MAF$, where $Sgn$ is randomly selected from $\left\{-1, 1\right\}$ and $MAF$ represents the corresponding minor allele frequency (MAF) at the signal. We begin with $\delta = 0.1$, then strengthen the signals by setting $\delta = 0.2$ to $0.7$. To demonstrate the empirical FWER, we also set $\delta = 0$. Since LRS has been shown not suitable for general signal region detection problems, and given that the only difference between Q-SCAN and LRS is the test statistic, we do not include LRS for comparison in the genetic setting. We compare the methods based on $1000$ Monte Carlo runs for computing FWERs and $500$ Monte Carlo runs over different $\delta$ for computing FDRs and TPRs. The FWERs of all settings are presented in Table \ref{table:FWER-Genetic}, and the FDRs and TPRs are in Figure \ref{fig:simulation_genetic}. Since SCANG-STAAR-O and SCANG-STAAR-B perform worse than SCANG-STAAR-S, we do not include them here for better visibility. The results for all three SCANG-STAAR methods are provided in Table S5 and S6 in the Supplementary Material.

\begin{table}
  \setlength{\abovecaptionskip}{0cm}
  \caption{\label{table:FWER-Genetic}\small FWERs of BiRS-DCF, BiRS-CL, Q-SCAN, 4S and SCANG-STAAR-S under genetic setting. The significant level is 0.05.} 
  \setlength{\tabcolsep}{4mm}
  \renewcommand\arraystretch{0.5}
  \centering
    \begin{tabular}{c|ccccc}
    \toprule
      & BiRS-DCF & BiRS-CL & Q-SCAN & SCANG-STAAR-S & 4S \\
    \midrule
     size & 0.0481 & 0.0442 & 0.0321 & 0.0402 & 0.0120 \\
    \bottomrule
    \end{tabular}
\end{table}

From the results, we see that the BiRS-DCF, BiRS-CL and SCANG-STAAR-S control the FWER, while the FWERs of Q-SCAN and 4S tend to be conservative. When the signal strength $\delta < 0.5$, BiRS-DCF achieves the highest TPRs with well-controlled FDRs. Although BiRS-CL, KnockoffScreen and SCANG-STAAR-S show higher TPRs than BiRS-DCF when $\delta \geq 0.5$, BiRS-CL exhibits elevated FDRs across among all $\delta$ values and KnockoffScreen shows significant FDR inflation when $\delta \geq 0.5$. For SCANG-STAAR-S, compared to Q-SCAN, the results indicate that the functional annotations indeed enhance detection accuracy. The 4S algorithm can only detect small short segments around some signal points and the TPRs are less than $0.4$. Based on these results, we conclude that the BiRS-DCF method performs well in terms of both FDR control and TPR across different settings.


We close this section by comparing the computational times under the genetic setting with $\delta = 0.7$ using 500 Monte Carlo runs, where the CPU employed is Xeon Gold 6448H. The average computational times are present in Table \ref{table:Computation}. showing that BiRS-DCF is more computationally efficient than Q-SCAN, KnockoffScreen and SCANG-STAAR-S. The 4S algorithm has a similar computational time but poorer detection accuracy. Moreover, for a direct comparison, we generate the DCF summary statistics first and only compare the computational time of scan and BiRS algorithm. The average detection time for BiRS is 1.77 minutes and the time for scan is 24.11 minutes, which is to some extent in line with the computational comparison in Section \ref{subsec:computational}.

\begin{table}
  \setlength{\abovecaptionskip}{0cm}
  \caption{\label{table:Computation}\small Computational times for BiRS-DCF, Q-SCAN, KnockoffScreen, SCANG-STAAR-S and 4S in genetic setting.} 
  \setlength{\tabcolsep}{2.5mm}
  \renewcommand\arraystretch{0.5}
  \centering
    \begin{tabular}{c|ccccc}
    \toprule
      & BiRS-DCF & Q-SCAN & KnockoffScreen & SCANG-STAAR-S & 4S \\
    \midrule
     Time(min) & 20.61 & 39.33 & 113.2 & 220.1 & 20.02 \\
    \bottomrule
    \end{tabular}
\end{table}


\section{Application to UK Biobank Data}
\label{sec:realdata}
We apply BiRS with the DCF test (BiRS-DCF) to perform whole-genome association studies for breast cancer, using individual-level data from women of European ancestry obtained from the UK Biobank (\url{https://biobank.ctsu.ox.ac.uk/crystal/index.cgi}). We first apply BiRS-DCF to conduct GWASs using genotype data extracted from UK Biobank field 22418. We then apply BiRS-DCF to analyze UK Biobank whole-genome sequencing (WGS) data from 200,000 individuals, the latest and largest dataset on actual usage for analyzing both coding and non-coding genomes \citep{Ribeiro2023, Wilcox2024.07.03.24309763}. To our knowledge, this is the first attempt to analyze the associations in the intergenic regions in real data analysis at a scale of hundreds of thousands of individuals. The phenotype for this study is based on UK Biobank field 40006, which corresponds to C50: Malignant neoplasm of the breast. Cases are defined as females with at least one record of ICD-10 code C50. We mention that the association studies are conducted without functional annotations, so we do not include SCANG-STAAR in the application.




\subsection{Quality control}
The sample exclusion criteria for quality control (QC) are as follows: we exclude individuals who were identified by UK Biobank as outliers based on either genotyping missingness rate or heterogeneity, whose sex inferred from the genotypes did not match their self-reported sex. We use the PCA analysis for population stratification extracted from UK Biobank and exclude individuals identified as non-Europeans. Specifically, we remove individuals whose one of the first two principal components deviated from the mean by five standard deviations. In addition, we delete individuals whose self-reported ethnicity was not of Europeans. We excluded individuals over 50 years of age to minimize the influence of age-related factors. Finally, we exclude individuals with a missingness greater than $5\%$ across variants which passed the UK Biobank quality control procedure. 

For the variants control, we only retain biallelic autosomal variants that were assayed by both genotyping arrays employed by UK Biobank. Variants who had failed UK Biobank quality control procedures in any of the genotyping batches are further excluded. Additionally, we remove variants with $P < 10^{-50}$ for departure from Hardy-Weinberg equilibrium (HWE) 
For the GWAS analysis, we exclude variants with a minor allele frequency (MAF) of less than $10^{-4}$. After quality control, the dataset includes 16,233 cases, 232,515 controls, and 604,745 variants across 22 chromosomes. In the WGS analysis, we only exclude variants with a minor allele count (MAC) of $\leq 1$. This leaves 6,779 cases, 96,356 controls, and 143,261,888 variants across 22 chromosomes in the 200k WGS dataset.

\subsection{Size evaluation}
To access the FWER control, we conduct permutation tests based on genotyped variants on each chromosome.  For chromosome $j$ $(j = 1, 2, \dots, 22)$,  in order to maintain the sample and variants ratio, we randomly select $\lfloor 16233*r_j \rfloor$ cases and $\lfloor 232515*r_j \rfloor$ controls, where $r_j = p_j/604745$ and $p_j$ is the number of variants in chromosome $j$. We permute cases and controls 1000 times, then apply the BiRS-DCF test on the permutation samples.  For comparison, we also implement BiRS with CL test (BiRS-CL) and Q-SCAN method. Q-SCAN was conducted using the summary statistics calculated from the generalized linear model, in which the covariates adjusted in the model included ages, assessment centers and leading 20 genomic principal components computed by UK Biobank. We calculate the family-wise error rates by determining the proportion of 1000 replications where at least one region was wrongly identified. Recall that KnockoffScreen is designed for FDR control (not suitable to be assessed by FWER), moreover, the computation for conducting permutation test by KnockoffScreen is prohibitively slow. For the same reason mentioned in simulation studies, the 4S coupled with DCF test is expected to be more conservative than BiRS-DCF. Thus we skip permutation test evaluations for KnockoffScreen and 4S, and apply them to identify the significant SNPs in the genotype data for detection comparison. 

The empirical FWERs of the three methods on all 22 chromosomes are presented in Table S7 of Supplementary Material. BiRS-CL obtains too much inflation (more than 35\% FWERs to detect associated SNPs at the significant level of 0.05).  BiRS-DCF controls the FWERs around the nominal level of 0.05 on all 22 chromosomes, and Q-SCAN is slightly inflated (more than 7\% FWERs) on chromosomes 15, 18, 19 and 22. As BiRS-CL shows inflation in the real data-based size evaluation, we only perform GWAS for C50 Malignant neoplasm of the breast using BiRS-DCF, Q-SCAN, KnockoffScreen and 4S. Moreover, we use REGENIE \citep{mbatchou2021computationally} as a benchmark that GWAS method established through single-SNP analysis.

\subsection{Genome-wide association studies (GWAS)}
\begin{figure}[!htbp]
\centering\includegraphics[width = 13cm]{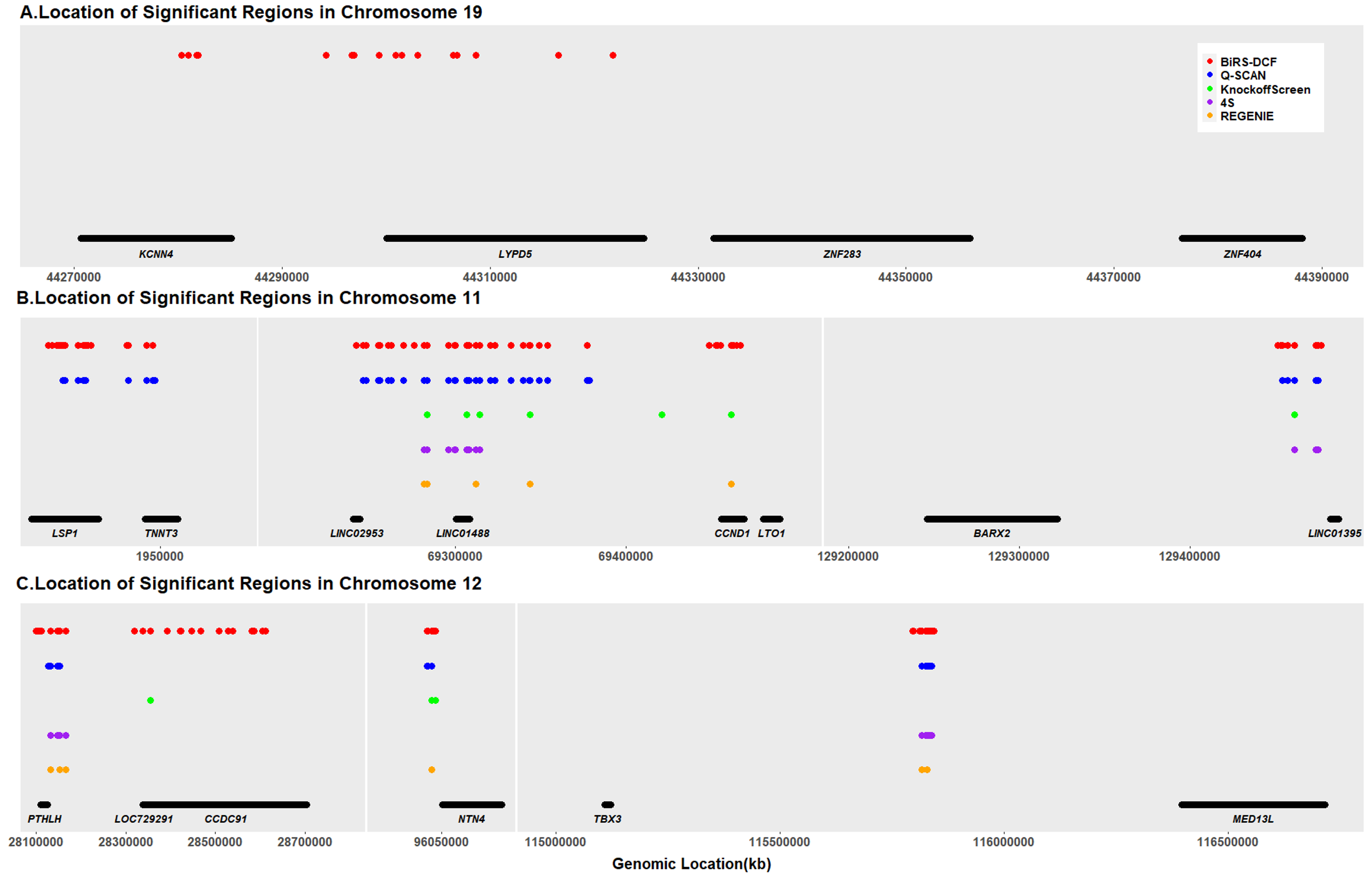}
\caption{\small Genetic landscape of the windows that were significantly associated with C50 Malignant neoplasm of breast on chromosome 19 (Panel A), 11 (Panel B) and 12 (Panel C) among Europeans in the UK Biobank GWASs. The physical positions of windows are based on build hg19.}
\label{fig:chr}
\end{figure}

Overall, BiRS-DCF identifies 552 SNPs associated with breast cancer. By comparison, Q-SCAN, KnockoffScreen, 4S and REGENIE identify 510, 93, 203 and 92 SNPs, respectively. The complete associated SNPs for these methods are listed in GWAS-BiRS.csv, GWAS-Scan.csv, GWAS-KS.csv, GWAS-4S.csv and GWAS-REGENIE.csv at
\url{https://github.com/ZWCR7/Supplementary-Material-for-BiRS.git}. Figure \ref{fig:chr} shows the signal regions detected by BiRS-DCF,  Q-SCAN,  KnockoffScreen, 4S and REGENIE on chromosome 11, chromosome 12 and 19. 

\textit{BiRS-DCF identifies new genome-wide significant regions on chromosome 19 (Panel A of Figure \ref{fig:chr}), encompassing related genes and SNPs that exhibits only weak associations in previous studies, where Q-SCAN, KnockoffScreen, 4S and REGENIE show no signals.} Specifically, signal regions are located in a 100kb length region from 4428000 to 4438000 bp in human genome reference assembly GRCh37, and mainly reside in genes \textsl{KCNN4} and \textsl{LYPD5}. \textsl{KCNN4} has been shown to regulate the proliferation of breast cancer cells and multidrug chemoresistance in breast cancer treatment \citep{lin2020kcnn4}. The region we detect within \textsl{LYPD5} covers two SNPs, rs349048 and rs11547806, which are potentially related to C50 Malignant neoplasm of breast from previous GWASs with $p$-value 0.006558 and 0.02365, respectively \citep{watanabe2019global}. In addition, both SNPs are significant expression quantitative trait loci for the gene \textsl{ZNF404} in the Beast Mammary tissue \citep{gtex2015genotype}, and it has been shown that \textsl{ZNF404} could lead to the development of breast cancer through mutations or altering the structural or functional activities \cite{masoodi2017computational}.  BiRS-DCF suggests that both SNPs rs349048 and rs11547806 are significantly associated with C50 Malignant neoplasm of breast, and further investigations are required to determine whether the GWAS associations are mediated through gene expression of \textsl{ZNF404}.

\textit{BiRS-DCF detects susceptible regions discussed by previous studies and accurately pinpoints SNPs that are associated with the risk of breast cancer in numerous studies, which are not identified by Q-SCAN, KnockoffScreen, 4S and REGENIE.} For instance, BiRS-DCF identifies an additional signal region near and within \textsl{CCND1} on chromosome 11, which is entirely missed by Q-SCAN and 4S.  \textsl{CCND1} is a candidate oncogene gene that encodes for D1 cyclin \citep{janssen2002myeov}.  D1 cyclin has been shown to regulate the progress of the cell cycle, and amplification and over-expression of \textsl{CCND1} are often observed in breast carcinoma \citep{janssen2002myeov}. Additionally,  the genes \textsl{LSP1} and  \textsl{TNNT3}, pinpointed by BiRS-DCF, are determined to be the susceptible loci for breast cancer based on previous studies. The variant rs3817198 located in LSP1 is recognized as a genome-wide significant SNP associated with the risk of breast cancer \citep{nourolahzadeh2020correlation,afzaljavan2020predictive}. BiRS-DCF detects rs3817198 with other 15 SNPs within \textsl{LSP1}, whereas other methods fail to detect SNP rs3817198. Furthermore, BiRS-DCF identifies a region near and within \textsl{LINC01488} with the highest number of associated SNPs compared to all other methods. \textsl{LINC01488} is a non-coding RNA regulated by estrogen, which regulates mammary gland development and is one of the main risk factors for breast cancer \citep{fanfani2021dissecting}.
Our results indicate that BiRS-DCF has enhanced power to detect signals more accurately.

\textit{BiRS-DCF outperforms other methods in detecting more significant regions, implying a potential interrelation between genes supported by biological evidence.} For example,  BiRS-DCF detects a signal region located about 18-90 kb upstream from the gene \textsl{CCDC91} on chromosome 12, and regions within \textsl{CCDC91}. \textsl{CCDC91} encodes an accessory protein that enables the protein binding activity and promotes the transportation of secreted proteins between Golgi and lysosomes \citep{mardones2007trans}.  It has been proposed that \textsl{CCDC91} is altered in breast tumors co-occurrences with \textsl{PTHLH}, and a possible interrelation between \textsl{PTHLH} and \textsl{CCDC91} might contribute to breast cancer susceptibility \citep{zeng2016identification}.  Consistently,  BiRS-DCF also detects multiple significant SNPs around \textsl{PTHLH},  suggesting that both \textsl{PTHLH} and \textsl{CCDC91} are target genes strongly associated with the risk of C50 Malignant neoplasm of breast. These signal regions cover several SNPs that are associated with C50 malignant neoplasm of the breast with marginal $p$-values smaller than $10^{-4}$ from previous GWASs. However, these $p$-values do not reach the genome-wide significance level of $5\times 10^{-8}$ \citep{watanabe2019global}. By comparison, Q-SCAN and REGENIE detect much fewer SNPs around \textsl{PTHLH}, and all other four methods almost miss the region around \textsl{CCDC91}. Our results demonstrate that BiRS-DCF is more powerful to detect important signal regions than these existing methods.

\subsection{Whole-genome sequencing (WGS) studies}
We apply BiRS-DCF to conduct a whole-genome scan using a 200,000-participant whole-genome sequencing (WGS) dataset from the UK Biobank Research Analysis Platform (RAP). The complete list of detected signal regions is available in BiRS-WGS-Region.csv at \url{https://github.com/ZWCR7/Supplementary-Material-for-BiRS.git}. Our analysis covers both the coding and non-coding regions of the genome. The coding regions contain DNA sequences that are transcribed into mRNA and translated into proteins, which perform essential cellular functions. In the non-coding regions, we assess the role of rare variants in the 5' and 3' untranslated regions (UTRs) and promoter regions obtained from the UCSC Genome Browser \citep{raney2024ucsc}. UTRs are segments located before (5' UTR) and after (3' UTR) the gene’s coding sequences that are transcribed but not translated into proteins. Promoters, on the other hand, are upstream of genes and serve as binding sites for proteins that initiate transcription. While these variants do not alter protein-coding sequences, they can affect the regulation of gene expression.

\textit{Several regions identified in the WGS analysis overlapped with our GWAS findings, including the susceptibility genes NEK10, SLC4A7, MRPS30, CCDC170, and ESR1, while we also uncover numerous rare variants within these genes that are underpowered in previous studies.} 
 On Chromosome 3, we detect 412 rare variants in \textsl{NEK10} with 9 rare variants in its 3’ UTRs regions. Previous burden testing suggests that \textsl{NEK10} may contribute to breast cancer risk through functional missense variants (P $\leq$ 0.0001), although none of the associations meet the burden testing significance thresholds \citep{decker2019targeted}. Notably, our method identifies rare variants that may occur within the conserved kinase domain, a region linked to breast cancer risk through functional alterations \citep{decker2019targeted}. 
 \textsl{NEK10} is highly expressed in normal breast tissue and likely acts as a tumor suppressor by limiting abnormal cell proliferation \citep{scannell2018germline}. Our identification of 9 rare 3' UTR variants underscores its possible role in post-transcriptional regulation during breast cancer progression. In addition, we confirm not only the known common variant rs4973768 associated with breast cancer risk in the \textsl{SLC4A7/NEK10} region, but also identify 23 rare variants in the 5' UTR of \textsl{SLC4A7}, and 65 more rare variants in its protein-coding region. \textsl{SLC4A7} regulates intracellular pH by mediating $Na+/HCO3^{-}$ cotransport in breast cancer cells, which plays a significant role in tumor progression \citep{axelsen2024antibodies}. Our discoveries position \textsl{NEK10} and \textsl{SLC4A7} as critical players in breast cancer susceptibility, with potential implications for the development of targeted therapies.

\textit{We uncover novel rare variants that provide fresh insights into previously uncharacterized regulatory and functional mechanisms.} On Chromosome 5, we detect a large number of rare variants in \textsl{MRPS30}, \textsl{MAP3K1}, \textsl{SETD9}, and \textsl{MIER3}. While \textsl{MRPS30} is identified in our GWAS analysis, the others are identified only through WGS analysis. Specifically, \textsl{MRPS30} plays a critical role in apoptosis, with its expression regulated by variants at the 5p12 locus, particularly SNP rs10941679 \citep{ghoussaini2016evidence}. Functional assays have demonstrated that SNP rs10941679 maps to an enhancer element that interacts with the \textsl{MRPS30} promoter region in breast cancer cell lines \citep{ghoussaini2016evidence}. We identify 19 rare variants in the promoter region of \textsl{MRPS30}, suggesting a potential contribution to altered gene expression and increased cancer susceptibility. Moreover, we detect 207 and 52 rare variants in the coding regions of \textsl{MAP3K1} and \textsl{SETD9}, respectively, along with an additional 8 rare variants in the 3' UTR of \textsl{MAP3K1}. \textsl{MAP3K1} contains SNP variants linked to breast cancer risk and mutations in \textsl{PIK3CA}. It has been observed that both \textsl{MAP3K1} and \textsl{SETD9} are overexpressed in tumors with \textsl{PIK3CA} mutations \citep{puzone2017snp}. Our identified rare variants suggest a possible synergistic effect between SNPs in the \textsl{MAP3K1/SETD9} region and \textsl{PIK3CA} mutations. Additionally, we identify 8 rare variants in the promoter region of \textsl{MIER3} and 16 rare variants in the 3' UTR of \textsl{CCDC170} on Chromosome 6. While both genes are linked to breast cancer through distinct mechanisms \citep{peng2017mier3,li2020therapeutic,jeong2021elucidation}, our findings highlight the crucial role of rare regulatory variants in contributing to disease progression. These results offer new insights and support the integration of these variants with GWAS data to enhance cancer risk prediction and inform more effective treatment strategies.

\section{Discussion}
\label{sec:dis}
In this work, we propose a novel binary and re-search (BiRS) algorithm to detect the existence and locations of signal regions. We show that the BiRS algorithm coupled with the DCF test (BiRS-DCF) controls the family-wise error rate well and guarantees the detection consistency under the framework of high-dimensional two-sample test. We demonstrate the efficiency of BiRS algorithm both empirically and theoretically by comparison with the state-of-art algorithms, including LRS, 4S, Q-SCAN, KnockoffScreen and SCANG-STAAR. Our simulation studies suggest that BiRS-DCF has enhanced detection accuracy under different data distributions, especially the genotype data for genetic association studies.   
The BiRS-DCF test identifies several new genetic regions for breast cancer without missing the former findings using the UK Biobank genomic data, which demonstrates that BiRS-DCF detects the locations and the sizes of signal regions more powerfully and accurately. Moreover, the BiRS-DCF test is powerful for detecting both coding and non-coding rare variant associations, including those in intergenic and regulatory regions, which provides new insights into disease mechanisms and targets for therapeutic intervention.

We make the following concluding remarks by pointing out some potential extensions that deserve future investigation. Firstly, incorporating functional annotation that provides valuable biological information can effectively increase power in rare variant association analysis \citep{Li2020dynamic}. Our framework is also flexible to incorporate various functional annotations by adding weights into features, where weights can be estimated using annotation principal components \citep{zhou2023favor}, CADD \citep{kircher2014general} or other methods. Secondly, we recognize the limitation of our two-sample framework in testing non-binary outcomes and correcting for complex covariates. Our application controls for key confounding factors (e.g., age, population stratification) by focusing on independent female European samples and implementing rigorous quality control procedures to ensure a more homogeneous population. To more effectively account for complex covariates, accommodate diverse populations and address non-binary traits, our future direction is to integrate the BiRS algorithm with various testing methods across different models, such as GLM in common regression analyses.

\bigskip
\begin{center}
{\large\bf SUPPLEMENTARY MATERIAL}
\end{center}


\begin{description}
\item[BiRS\_supp]  Theoretical properties, technical assumptions, lemmas, and proofs of the theoretical results, along with additional simulation and application results for ``Fast Signal Region Detection with Application to Whole Genome Association Studies''. (.pdf file)
\item[Code and Data for BiRS] R-package {\tt BiRS} of the algorithm, the simulation and application codes, as well as the GWASs and WGS results at\\ 
\url{https://github.com/ZWCR7/Supplementary-Material-for-BiRS.git}.

\end{description}
\small
\bibliographystyle{agsm}

\bibliography{Bibliography-MM-MC}
\end{document}